\documentclass{article}
 \input{epsf}
\usepackage{amsmath}
\usepackage{graphicx}
\usepackage{verbatim}
\usepackage{epsf}
\usepackage{latexsym}
\usepackage{amsmath,amsfonts,amssymb,amsthm}

\newcommand{\be}{\begin{equation}}
\newcommand{\ee}{\end{equation}}
\newcommand{\beq}{\begin{equation}}
\newcommand{\eeq}{\end{equation}}
\newcommand{\ba}{\begin{array}}
\newcommand{\ea}{\end{array}}
\newcommand{\bi}{\begin{itemize}}
\newcommand{\ei}{\end{itemize}}
\newcommand{\bea}{\begin{eqnarray}}
\newcommand{\eea}{\end{eqnarray}}
\newcommand{\ben}{\begin{enumerate}}
\newcommand{\een}{\end{enumerate}}
\newcommand{\bean}{\begin{eqnarray*}}
\newcommand{\eean}{\end{eqnarray*}}
\newcommand{\eref}[1]{(\ref{#1})}

\newcommand{\nn}{\nonumber}

\numberwithin{equation}{section}

\begin{document}

\title{The covariant perturbative
string spectrum}

\author{Davide Forcella, Amihay Hanany and Jan Troost}

\maketitle

\begin{center}

Theoretical Physics Group, The Blackett Laboratory \\
Imperial College London, Prince Consort Road\\ 
London,  SW7 2AZ,  UK \\
and \\
 Laboratoire de Physique Th\'eorique\footnote{Unit\'e Mixte du CNRS et
    de l'Ecole Normale Sup\'erieure associ\'ee \`a l'universit\'e Pierre et
    Marie Curie 6, UMR
    8549. Preprint: LPTENS-10/28, Imperial/TP/10/AH/04.} \\
Ecole Normale Sup\'erieure  \\
24 rue Lhomond \\ F--75231 Paris Cedex 05 \\ France
\end{center}

\begin{abstract}We provide generating functions for the perturbative massive string
  spectrum which are covariant with respect to the $SO(9)$ little
  group, and which contain all the representation theoretic content of
  the spectrum. Generating functions for perturbative bosonic, Type II,
  Heterotic and Type I string theories are presented, and generalizations
  are discussed.
\end{abstract}

%\preprint{Imperial/TP/10/AH/04\\
%LPTENS-10/28}

\tableofcontents

\section{Introduction}
The multiplicities of the perturbative on-shell string states as a
function of their mass is known, for instance through the light-cone
gauge
partition function.  For a string theory in $(D-1,1)$ Minkowksi space,
the expression can easily be made $SO(D-2)$ covariant. Massive string
states however are classified on-shell by the little group
$SO(D-1)$. Although it is known that the perturbative string spectrum
respects this symmetry since it respects Lorentz invariance, no
generating functions for the $SO(D-1)$ representation content of string
states at all mass levels has been given. It is our purpose in this
paper to provide such generating functions.

\section{Two tools}
In order to write down the generating functions in a compact form,
we introduce two tools. Firstly, we discuss
characters for irreducible representations characterized by their Dynkin
labels. Secondly, we introduce plethystics.
\subsection{Characters and Dynkin labels}
An irreducible representation of the rotation group $SO(n)$ is
characterized by its Dynkin labels, $a_1, a_2, \ldots, a_r$, where $r$
is the rank of $SO(n)$ and the $a_i$ are non-negative integers. The
integers give the coefficient of the fundamental weights in the
highest weight of the irreducible representation of $SO(n)$. We denote
the corresponding character by $[a_1, a_2, \ldots, a_r]_n$. For the
purpose of this note, the main groups of interest are $SO(8)$ and
$SO(9)$. In ten dimensions,
these are the little groups for massless and
massive particles  respectively. Each
representation of these groups carries four Dynkin labels.
  To each fundamental
weight we assign a fugacity. Each weight in the irreducible
representation gives rise to a term in the character formula which
raises the fugacity to the power equal to the coefficient of the
fundamental weight in the weight. For $SO(8)$ we denote the four
fugacities $z_i$ and for $SO(9)$ we denote them $y_i$.

Explicit formulae are needed for the characters of the
eight-dimensional vector representation with character $[1,0,0,0]_8$,
the eight-dimensional spinor representation $[0,0,1,0]_8$ and the
eight-dimensional complex conjugate spinor representation
$[0,0,0,1]_8$:
\begin{eqnarray}
\label{so8reps}
\left[1,0,0,0 \right]_8 &=& z_1+\frac{z_2}{z_1}+\frac{z_3 z_4}{z_2}+\frac{z_4}{z_3}+\frac{z_3}{z_4}+\frac{z_2}{z_3 z_4}+\frac{z_1}{z_2}+\frac{1}{z_1}, \nonumber \\
\left[0,0,1,0\right]_8 &=& z_3+\frac{z_2}{z_3}+\frac{z_4 z_1}{z_2}+\frac{z_1}{z_4}+\frac{z_4}{z_1}+\frac{z_2}{z_4 z_1}+\frac{z_3}{z_2}+\frac{1}{z_3}, \nonumber \\
\left[0,0,0,1\right]_8 &=& z_4+\frac{z_2}{z_4}+\frac{z_1 z_3}{z_2}+\frac{z_3}{z_1}+\frac{z_1}{z_3}+\frac{z_2}{z_1 z_3}+\frac{z_4}{z_2}+\frac{1}{z_4}.
\end{eqnarray}
Note that these formulae reflect the $SO(8)$ triality which act by permuting the Dynkin labels $a_1, a_3$, and $a_4$ in addition to the fugacities, $z_1, z_3$, and $z_4$.
% For reference, the dimension formula for an $SO(8)$ representation is given by a 12 order polynomial in $a_i$,
% \bea
% &&\dim [a_1, a_2, a_3, a_4]_8 = 
% \frac{1}{4320}(a_1+1) (a_2+1) (a_3+1) (a_4+1) (a_1+a_2+2) \nonumber \\
% && (a_2+a_3+2) (a_2+a_4+2) (a_1+a_2+a_3+3) (a_1+a_2+a_4+3) \nonumber \\
% && (a_2+a_3+a_4+3) (a_1+a_2+a_3+a_4+4) (a_1+2 a_2+a_3+a_4+5) .
% \eea

We also give the 
$SO(9)$ characters corresponding to the nine-dimensional 
vector representation $[1,0,0,0]_9$, and the sixteen-dimensional 
spinor representation $[0,0,0,1]_9$:
\begin{eqnarray}
\left[1,0,0,0 \right]_9 &=& y_1+\frac{y_2}{y_1}+\frac{y_3}{y_2}+\frac{y_4^2}{y_3}+\frac{y_3}{y_4^2}+\frac{y_2}{y_3}+\frac{y_1}{y_2}+\frac{1}{y_1}+1, \\
\left[0,0,0,1\right]_9 &=& \frac{y_3}{y_4}+\frac{y_2 y_4}{y_3}+\frac{y_4 y_1}{y_2}+\frac{y_1}{y_4}+\frac{y_4}{y_1}+\frac{y_2}{y_4 y_1}+\frac{y_3}{y_2 y_4}+\frac{y_4}{y_3} \nonumber \\ \label{reps}
&+&y_4+\frac{y_2}{y_4}+\frac{y_1 y_3}{y_4 y_2}+\frac{y_3}{y_1 y_4}+\frac{y_1 y_4}{y_3}+\frac{y_4 y_2}{y_1 y_3}+\frac{y_4}{y_2}+\frac{1}{y_4}.
\end{eqnarray}
% For reference, the dimension formula for an $SO(9)$ representation is given by a 16 order polynomial in $a_i$,
% \bea
% &&\dim [a_1, a_2, a_3, a_4]_9 = \frac{1}{3628800}(a_1+1) (a_2+1) (a_3+1) (a_4+1) (a_1+a_2+2) (a_2+a_3+2) \nonumber \\
% && (a_3+a_4+2) (a_1+a_2+a_3+3) (a_2+a_3+a_4+3) (2 a_3+a_4+3) (a_1+a_2+a_3+a_4+4) \nonumber \\
% && (a_2+2 a_3+a_4+4) (a_1+a_2+2 a_3+a_4+5) (2 a_2+2 a_3+a_4+5) (a_1+2 a_2+2 a_3+a_4+6) \nonumber \\ 
% && (2 a_1+2 a_2+2 a_3+a_4+7) .
% \eea
The decomposition of $SO(9)$ representations into irreducible
representations of $SO(8)$ can be read off from the characters
by relating the $SO(9)$ fugacities to those of $SO(8)$:
\begin{equation}
\label{so9s08dec}
y_1 = z_1, \qquad y_2 = z_2, \qquad y_3 = z_3 z_4, \qquad y_4 = z_4.
\end{equation}
Thus, using \eref{so9s08dec}, the first line in equation \eref{reps}
for $[0,0,0,1]_9$, the character for the spinor representation of
$SO(9)$, can be seen to correspond to the $[0,0,1,0]_8$ spinor representation
and the second line in equation \eref{reps} corresponds to the conjugate
spinor
$[0,0,0,1]_8$. Note that generically, the reconstruction of
$SO(9)$ representations from their $SO(8)$ reductions is ambiguous.
Nevertheless, for the low-dimensional representations discussed above,
it is clear that the $SO(8)$ weights lift uniquely 
to $SO(9)$ weights as follows:
\begin{eqnarray}
\left[1,0,0,0 \right]_8 &=& \left[1,0,0,0 \right]_9 - 1, \nonumber \\
\left[0,0,1,0\right]_8 &=& \frac{y_3}{y_4}+\frac{y_2 y_4}{y_3}+\frac{y_4 y_1}{y_2}+\frac{y_1}{y_4}+\frac{y_4}{y_1}+\frac{y_2}{y_4 y_1}+\frac{y_3}{y_2 y_4}+\frac{y_4}{y_3}, \nonumber \\
\left[0,0,0,1\right]_8 &=& y_4+\frac{y_2}{y_4}+\frac{y_1 y_3}{y_4 y_2}+\frac{y_3}{y_1 y_4}+\frac{y_1 y_4}{y_3}+\frac{y_4 y_2}{y_1 y_3}+\frac{y_4}{y_2}+\frac{1}{y_4}.
\label{so8}
\end{eqnarray}

\subsection{Plethystics}
Our second tool will be to rewrite various combinatorial expressions in terms of the
formalism of plethystics. We collect here the definition of various plethystic functions.
Some applications of
plethystic functions to problems in string theory and
supersymmetric gauge theory can be found in \cite{pleth} and
references thereto.

For a function of $m$ variables $g(t_1, \ldots, t_m)$ that vanishes at
the origin, $g(0, \ldots, 0) =0$, the plethystic exponential is
defined to be \beq PE \left [ g(t_1, \ldots, t_m) \right ] =
\exp\left(\sum_{k=1}^\infty \frac{g(t_1^k, \ldots, t_m^k)}{k} \right
).
\label{PEB}
\eeq The fermionic plethystic exponential contains extra minus signs:
\beq PE_F \left [ g(t_1, \ldots, t_m) \right ] =
\exp\left(\sum_{k=1}^\infty \frac{(-1)^{k+1} g(t_1^k, \ldots,
   t_m^k)}{k} \right ).
\label{PEF}
\eeq The inverse of the plethystic exponential is called the
plethystic logarithm and is defined for a function of $m$ variables
$g(t_1, \ldots, t_m)$ that is equal to 1 at the origin, $g(0, \ldots,
0) =1$, as:
\beq PL \left [ g(t_1, \ldots, t_m) \right ] =
\sum_{k=1}^\infty \frac{\mu(k) \log g(t_1^k, \ldots, t_m^k)}{k}, \eeq
with $\mu(k)$ the M\"obius function, \beq \mu(k) =
\begin{cases}
(-1)^n & \text{$k$ is a product of $n$ distinct primes,} \\
0 & \text{otherwise.}
\end{cases} 
\eeq
We  now have the tools to tackle the generating functions.

%%%%%%%%%%%%%%%%%%%%%%%%%%%% 

\section{The covariant perturbative string
partition functions} \label{TypeII}
\subsection{The chiral ten-dimensional partition function}
Let us concentrate on the left-movers of a Type II string, and on the
integrand appearing in the partition function (i.e. the integral
stripped of both 
momentum zero-modes
and the integration of the modular parameter
over the fundamental domain). It is our goal
to render the integrand manifestly $SO(9)$ covariant at all massive levels.
\subsection*{The problem}
Before introducing fugacities, the partition function takes the form:
\begin{equation}
\label{partitionII}
Z(q) = 16 \prod_{n=1}^\infty \left (\frac{1+q^n}{1-q^n}\right)^8,
\end{equation}
where $q$ is the fugacity that counts the mass level of the
perturbative string spectrum.
After combining with the right-movers, the expansion of this function gives
the number of physical polarization modes at any given mass.
The first few terms in the expansion
\begin{equation}
Z(q) = 16 + 256q + 2304 q^2 + 15360 q^3 + 84224 q^4 + O(q^5),
\end{equation}
give information about the massless and massive spectrum of the open
and Type II string.  At zeroth order we find the 16 polarization modes
of the massless vector multiplet in 9+1 dimensions, which decompose
under $SO(8)$ as the vector representation $[1,0,0,0]_8$ and the
spinor representation $[0,0,0,1]_8$, corresponding to the 9+1
dimensional gauge field and the 9+1 dimensional gaugino respectively:
\beq Z_0 = [1,0,0,0]_8+[0,0,0,1]_8.
\label{Z0}
\eeq At the first massive level we find the $SO(9)$ representations
which also appear in the massless supergravity multiplet in 10+1
dimensions: \beq Z_Q = [2,0,0,0]_9+[1,0,0,1]_9+[0,0,1,0]_9.
\label{ZQ}
\eeq This multiplet corresponds to a multiplet of supercharges
and encodes the supersymmetric nature of
all massive representations in ten dimensions. Any higher order massive
supermultiplet is a tensor product of this multiplet with another
representation of $SO(9)$. 
 We can therefore rewrite the partition
function in a factorized form
\begin{equation}
Z(q) = 16 + 256q Z_m(q); \qquad Z_m(q) = 1 + 9 q + 60 q^2 + 329 q^3 + O(q^4),
\label{Zq}
\end{equation}
where the first equation serves as the definition of $Z_m(q)$, the
partition function for perturbative massive modes, and the second
equation is the expansion of $Z_m(q)$ to first few orders.  Due to
this factorization we can proceed a little further in reconstructing
the irreducible representations of $SO(9)$ that appear in the
partition function, on the basis of their dimensions only.  At the
second mass level we find the representation content \beq \left(
  [2,0,0,0]_9+[1,0,0,1]_9+[0,0,1,0]_9\right) [1,0,0,0]_9, \eeq and at
the third mass level we get \beq \left(
  [2,0,0,0]_9+[1,0,0,1]_9+[0,0,1,0]_9\right) \left([2,0,0,0]_9 +
  [0,0,0,1]_9\right).  \eeq This information can also be found in text
books.  At low levels, the identification of $SO(9)$ representations
on the basis of their dimensions only
is unique. At higher mass
levels this is no longer true. 

\subsection*{The solution}
To gain further insight, we recall that the perturbative Type II
spectrum is made out of a tower of eight bosonic oscillators and eight
fermionic oscillators that transform as vectors under $SO(8)$. We can
therefore introduce the four fugacities of $SO(8)$ and refine the
partition function to include the characters of $SO(8)$ and not just
their dimensions. For the massive spectrum, we further wish to extend
these characters into $SO(9)$ characters as the massive on-shell
spectrum decomposes into irreducible representations of the little
group $SO(9)$. 

The eight
bosonic oscillators transform in the vector representation of $SO(8)$
and carry a level contribution $n$. The tower of string states is formed by
symmetrization of those oscillators. The plethystic exponential 
precisely keeps track of the symmetrization procedure.
The $SO(8)$ covariant bosonic part of the  partition function is therefore:
\beq Z_B (q; z_1, z_2, z_3, z_4) = PE \left[ \frac{q}{1-q}
 [1,0,0,0]_8 \right].
\label{ZB}
\eeq
If we set the $SO(8)$ fugacities to one, we recover the infinite denominator in equation \eref{partitionII}:
\beq
Z_B (q; 1, 1, 1, 1) = PE \left[ \frac{8 q}{1-q} \right] = \prod_{n=1}^\infty\frac{1}{(1-q^n)^8}.
\label{ZB1}
\eeq 
The (worldsheet) fermionic partition function is slightly more involved.
The antisymmetrization can be treated by the fermionic
plethystic exponential. To implement GSO, we also wish to keep track of the
fermion number of all states, for which we introduce an extra fugacity $f$:
\beq Z_F (q; f; z_1, z_2, z_3, z_4) = PE_F \left[
 \frac{f}{1-q} [1,0,0,0]_8 \right].
\label{ZF}
\eeq Again, after setting the $SO(8)$ fugacities to one we find the
infinite numerator in equation \eref{partitionII} (supplemented with
the fermion number fugacity): \beq Z_F (q; f; 1, 1, 1, 1) = PE_F
\left[ \frac{8 f}{1-q} \right] = \prod_{n=0}^\infty{(1+ f q^n)^8}.
\eeq 
In the formulas below, the $q$ and $z_i$ dependence is kept implicit. Only the
$f$ dependence is mentioned explicitly, namely $Z_F(f)$.
% Note that by the definitions of the plethystic
%exponentials \eref{PEB} and \eref{PEF} and the partition functions
%\eref{ZB} and \eref{ZF} we have the identity 
%\beq Z_B (q; z_1, z_2,
%z_3, z_4) Z_F (q; -q; z_1, z_2, z_3, z_4) = 1,
%\label{identity}
%\eeq which follows from worldsheet supersymmetry.  
We define the GSO projected partition functions in the Neveu-Schwarz
%
%In the NS sector the vacuum is
%a singlet of $SO(8)$ and fermions carry half integral level. The GSO
%projection restricts to only odd number of fermions. Therefore we
%define the NS partition function 
\beq Z_{NS} =
\frac{1}{2\sqrt{q}}\left(Z_F\left(\sqrt{q}\right)-Z_F\left(-\sqrt{q}\right)\right),
\label{ZNS}
\eeq
and in the Ramond-sector
%
%In the R sector the vacuum has a 16 fold degeneracy and it transforms
%as $[0,0,1,0]_8$ and $[0,0,0,1]_8$. On one set of vacua, the GSO
%projection restricts to an even number of oscillators acting and on the other
%an odd number of oscillators act. We therefore define even and odd
%functions of $f$, 
\beq Z_{\pm} =
\frac{1}{2}\left(Z_F\left({q}\right)\pm Z_F\left(-{q}\right)\right).
\label{Zpm}
\eeq With these definitions we can collect the different fermionic
partition functions into a single partition function that takes into
account the boundary conditions, the GSO projection, and
the fugacities associated to the vacua \beq Z_{RNS}
(q; z_1, z_2, z_3, z_4) = Z_{NS}+[0,0,0,1]_8 Z_{+} + [0,0,1,0]_8
Z_{-}.  \eeq
The open string, or left-moving partition function takes the form
\beq
\label{ZOpen}
Z_{Left} (q; z_1, z_2, z_3, z_4) = Z_B \left( Z_{NS}+[0,0,0,1]_8 Z_{+} + [0,0,1,0]_8 Z_{-} \right ).
\eeq
Equation \eref{ZOpen} gives the refined partition function of the open
string as a function of  $SO(8)$ fugacities. It remains to rewrite it in
terms of $SO(9)$ characters, once the massless sector is
subtracted. This turns out to be a simple task by realizing that there
are only 3 $SO(8)$ representations which are involved in computing
$Z_{Left}$. These are the vector and the two spinor
representations. Equations \eref{so8} provide the final ingredient, as
it rewrites the characters of these representations in terms of
$SO(9)$ fugacities. Thus using \eref{ZOpen}, \eref{ZB}, \eref{ZF},
\eref{ZNS}, and \eref{Zpm} we arrive at the open string partition
function written in terms of $SO(9)$ fugacities.
Using the equations for the characters
of the spinor representations of $SO(9)$ and $SO(8)$ we can rewrite equation \eref{ZOpen}:
%\beq
%Z_{Open} (q; z_1, z_2, z_3, z_4) = Z_B \left( Z_{NS}+\frac{1}{2}[0,0,0,1]_9 Z_{F}(q) +\frac{1}{2}\left( [0,0,1,0]_8 - [0,0,0,1]_8 \right) Z_{F} (-q) \right ).
%\eeq
%Using identity \eref{identity} we get a simplified expression,
\beq
Z_{Left} (q; z_1, z_2, z_3, z_4) = \frac{1}{2}\left( [0,0,0,1]_8 - [0,0,1,0]_8 \right) + Z_B \left( Z_{NS}+\frac{1}{2}[0,0,0,1]_9 Z_{F}(q)\right ) .
\label{ZOpen1}
\eeq We provide explicit expressions for the functions appearing in
the formula above: \bea
Z_B &=& PE \left[ \frac{q}{1-q} \left ([1,0,0,0]_9 - 1\right ) \right] ,\nn \\
Z_{NS} &=& \frac{1}{2\sqrt{q}}\left(PE_F \left[ \frac{\sqrt{q}}{1-q} \left ( [1,0,0,0]_9 - 1\right) \right]-PE_F \left[ \frac{(-\sqrt{q})}{1-q} \left ( [1,0,0,0]_9 - 1\right) \right] \right) ,\nn \\
Z_{F}(q) &=& PE_F \left[ \frac{q}{1-q} \left ( [1,0,0,0]_9 - 1\right)
\right] .  \eea The first term in the partition function
\eqref{ZOpen1} has no $q$ dependence and contributes only at the
massless level, while the second contribution is manifestly $SO(9)$
covariant.  The factor $1/2$ appears to contribute fractional
coefficients to irreducible representations. One can show that this
is not the case either by explicit expansion as
demonstrated in the next section or by the following general logic. Since the arguments of
$Z_B$ and $Z_F(q)$ are equal, the contribution to the plethystic
exponential gets a factor 2 from odd powers of the argument. This
factor 2 in the exponential therefore cancels the 1/2 order by order.  The
generating function \eqref{ZOpen1} for the full $SO(9)$ covariant
content of the perturbative string spectrum is our main result. Our
construction makes it very plausible that the generating function has
positive integer coefficients for each $SO(9)$ character at each mass
level. It is a challenge to prove this explicitly.

\subsection*{The Covariant Partition Function for Massive Modes}
With this result at hand, we can also refine the expression
for the massive spectrum, with factored massive supermultiplet.
We define:
\beq Z_m(q; y_1, y_2, y_3, y_4) = \frac{Z_{Left} - Z_0}{q Z_Q },
\eeq with $Z_{Left}$ and $Z_Q$ given in equations \eref{ZOpen1},
and \eref{ZQ} and $Z_0$ equal to the partition function of 
the massless modes.
The expansion of the function 
$Z_m$  in powers of $q$ and $SO(9)$ characters can be performed explicitly (using
a symbolic manipulation program) and gives
\bea
&&Z_m = 1 + [1,0,0,0]_9 q + \left([2,0,0,0]_9 + [0,0,0,1]_9\right) q^2 \nonumber \\
&+& ([3, 0, 0, 0]_9 + [1, 0, 0, 1]_9 +[1, 0, 0, 0]_9 + [0, 1, 0, 0]_9) q^3 \nonumber \\
&+& ([4, 0, 0, 0]_9 + [2, 0, 0, 1]_9 + [2, 0, 0, 0]_9 + [1, 1, 0, 0]_9 + [1, 0, 0, 1]_9 + [0, 1, 0, 0]_9 + [0, 0, 1, 0]_9 \nonumber \\
&+& [0, 0, 0, 1]_9 + [0,0,0,0]_9) q^4 \nonumber \\
&+& ([5, 0, 0, 0]_9 + [3, 0, 0, 1]_9 + [3, 0, 0, 0]_9 + [2, 1, 0, 0]_9 + [2, 0, 0, 1]_9 + [2, 0, 0, 0]_9 + 2 [1, 1, 0, 0]_9 \nonumber \\
&+& [1, 0, 1, 0]_9 + 2 [1, 0, 0, 1]_9 + 2 [1, 0, 0, 0]_9 + [0, 1, 0, 1]_9 + [0, 1, 0, 0]_9 + [0, 0, 0, 2]_9 + 2 [0, 0, 0, 1]_9) q^5 \nonumber \\
&+& ([6, 0, 0, 0]_9 + [4, 0, 0, 1]_9 + [4, 0, 0, 0]_9 + [3, 1, 0, 0]_9 + [3, 0, 0, 1]_9 + [3, 0, 0, 0]_9 + 2 [2, 1, 0, 0]_9 \nonumber \\
&+& [2, 0, 1, 0]_9 + 3 [2, 0, 0, 1]_9 + 3 [2, 0, 0, 0]_9 + [1, 1, 0, 1]_9 + 2 [1, 1, 0, 0]_9 + [1, 0, 1, 0]_9 + [1, 0, 0, 2]_9 \nonumber \\
&+& 4 [1, 0, 0, 1]_9 + 2 [1, 0, 0, 0]_9 + [0, 2, 0, 0]_9 + 2 [0, 1, 0, 1]_9 + 2 [0, 1, 0, 0]_9 + 3 [0, 0, 1, 0]_9 + [0, 0, 0, 2]_9 \nonumber \\
&+& 2 [0, 0, 0, 1]_9 + 2 [0, 0, 0, 0]_9) q^6
\nonumber \\
&+& ([7, 0, 0, 0]_9 + [5, 0, 0, 1]_9 + [5, 0, 0, 0]_9 + [4, 1, 0, 0]_9 + [4, 0, 0, 1]_9 + [4, 0, 0, 0]_9 + 2 [3, 1, 0, 0]_9
\nonumber \\
&+& [3, 0, 1, 0]_9 + 3 [3, 0, 0, 1]_9 + 4 [3, 0, 0, 0]_9 + [2, 1, 0, 1]_9 + 3 [2, 1, 0, 0]_9 + [2, 0, 1, 0]_9 + [2, 0, 0, 2]_9
\nonumber \\
&+& 5 [2, 0, 0, 1]_9 + 3 [2, 0, 0, 0]_9 + [1, 2, 0, 0]_9 + 3 [1, 1, 0, 1]_9 + 5 [1, 1, 0, 0]_9 + 4 [1, 0, 1, 0]_9 + 2 [1, 0, 0, 2]_9
\nonumber \\
&+& 7 [1, 0, 0, 1]_9+ 5 [1, 0, 0, 0]_9+ [0, 2, 0, 0]_9 + [0, 1, 1, 0]_9 + 4 [0, 1, 0, 1]_9 + 5 [0, 1, 0, 0]_9 + [0, 0, 1, 1]_9 \nonumber \\
&+& 2 [0, 0, 1, 0]_9 + 3 [0, 0, 0, 2]_9 + 4 [0, 0, 0, 1]_9 + [0, 0, 0, 0]_9 ) q^7
\nonumber \\
&+& ([8, 0, 0, 0]_9 + [6, 0, 0, 1]_9 + [6, 0, 0, 0]_9 + [5, 1, 0, 0]_9 + [5, 0, 0, 1]_9 + [5, 0, 0, 0]_9 + 2 [4, 1, 0, 0]_9
\nonumber \\
&+& [4, 0, 1, 0]_9 + 3 [4, 0, 0, 1]_9 + 4 [4, 0, 0, 0]_9 + [3, 1, 0, 1]_9 + 3 [3, 1, 0, 0]_9 + [3, 0, 1, 0]_9 + [3, 0, 0, 2]_9
\nonumber \\
&+& 6 [3, 0, 0, 1]_9 + 4 [3, 0, 0, 0]_9 + [2, 2, 0, 0]_9 + 3 [2, 1, 0, 1]_9 + 6 [2, 1, 0, 0]_9 + 5 [2, 0, 1, 0]_9 + 2 [2, 0, 0, 2]_9
\nonumber \\
&+& 10 [2, 0, 0, 1]_9 + 9 [2, 0, 0, 0]_9 +2[1,2,0,0]_9 + [1, 1, 1, 0]_9 + 6 [1, 1, 0, 1]_9 + 9 [1, 1, 0, 0]_9 + [1, 0, 1, 1]_9
\nonumber \\
&+& 6 [1, 0, 1, 0]_9 + 6 [1, 0, 0, 2]_9 + 12 [1, 0, 0, 1]_9 + 5 [1, 0, 0, 0]_9 + [0, 2, 0, 1]_9 + 4 [0, 2, 0, 0]_9 + [0, 1, 1, 0]_9
\nonumber \\
&+& [0, 1, 0, 2]_9 + 8 [0, 1, 0, 1]_9 + 7 [0, 1, 0, 0]_9 + 3 [0, 0, 1, 1]_9 + 7 [0, 0, 1, 0]_9 + 4 [0, 0, 0, 2]_9 + 8 [0, 0, 0, 1]_9
\nonumber \\
&+& 3 [0, 0, 0, 0]_9 ) q^8
%\nonumber
\\
&+& ([9, 0, 0, 0]_9 + [7, 0, 0, 1]_9 + [7,0,0,0]_9 + [6, 1, 0, 0]_9 + [6, 0, 0, 1]_9 + [6, 0, 0, 0]_9 + 2 [5, 1, 0, 0]_9 
\nonumber \\
&+& [5, 0, 1, 0]_9 + 3 [5, 0, 0, 1]_9 + 4 [5, 0, 0, 0]_9 + [4, 1, 0, 1]_9 + 3 [4, 1, 0, 0]_9 + [4, 0, 1, 0]_9 + [4, 0, 0, 2]_9 
\nonumber \\
&+& 6 [4, 0, 0, 1]_9 + 5 [4, 0, 0, 0]_9 + [3, 2, 0, 0]_9 + 3 [3, 1, 0, 1]_9 + 7 [3, 1, 0, 0]_9 + 5 [3, 0, 1, 0]_9 + 2 [3, 0, 0, 2]_9
\nonumber \\
&+& 11 [3, 0, 0, 1]_9 + 11 [3, 0, 0, 0]_9 + 2 [2, 2, 0, 0]_9 + [2, 1, 1, 0]_9 + 7 [2, 1, 0, 1]_9 + 12 [2, 1, 0, 0]_9 + [2, 0, 1, 1]_9
\nonumber \\
&+& 7 [2, 0, 1, 0]_9 + 7 [2, 0, 0, 2]_9 + 19 [2, 0, 0, 1]_9 + 10 [2, 0, 0, 0]_9 + [1, 2, 0, 1]_9 + 5 [1, 2, 0, 0]_9 + 2 [1, 1, 1, 0]_9
\nonumber \\
&+& [1, 1, 0, 2]_9 + 14 [1, 1, 0, 1]_9 + 17 [1, 1, 0, 0]_9 + 4 [1, 0, 1, 1]_9 + 15 [1, 0, 1, 0]_9 + 9 [1, 0, 0, 2]_9 + 22 [1, 0, 0, 1]_9
\nonumber \\
&+& 12 [1, 0, 0, 0]_9 + [0, 3, 0, 0]_9 + 2 [0, 2, 0, 1]_9 + 5 [0, 2, 0, 0]_9 + 6 [0, 1, 1, 0]_9 + 3 [0, 1, 0, 2]_9 + 15 [0, 1, 0, 1]_9
\nonumber \\
&+& 13 [0, 1, 0, 0]_9 + 5 [0, 0, 1, 1]_9
+ 10 [0, 0, 1, 0]_9 + 2 [0, 0, 0, 3]_9 + 10 [0, 0, 0, 2]_9 + 12 [0, 0, 0, 1]_9 + 3 [0, 0, 0, 0]_9 ) q^9
\nonumber \\
&+& O(q^{10}).
\nonumber
\label{pertZm}
\eea The coefficients are positive integers.  These results agree with
Appendix B of \cite{Bianchi:2003wx} where they were computed order by
order in terms of $SO(8)$ representations, and then reconstituted into
$SO(9)$ representations.  Using our generating formula, we easily
produce covariant results at higher orders, as demonstrated above.

We point out a selection rule which arises up to this
order in the mass expansion which states that the third entry in the
set of Dynkin labels is either 0 or 1 but not higher. It would be
interesting to attempt to prove such a selection rule for the full perturbative
spectrum.
\subsection*{Factoring symmetric tensors}
We note that $Z_m$ starts with 1, suggesting it is appropriate to take
a plethystic logarithm.  The result to order $q^2$ is \bea PL[Z_m] =
[1,0,0,0]_9 q + \left([0, 0, 0, 1]_9 - 1\right) q^2 + O(q^3).  \eea
This result implies that at order $q^n$ there is an $n$-th symmetric
product of the vector representation, which is a reducible
representation. Indeed, this pattern is observed in equation
\eref{pertZm}. In order to proceed let us take a small detour and
recall a general formula for completely symmetric tensor
representations of orthogonal groups. See \cite{Hanany:2008qc} for a
related discussion.  The symmetrization of the vector representation
is naturally given by taking the plethystic exponential and it
satisfies the following identity 
\bea
PE \left [ [1,0, \ldots, 0]_n q \right] 
&=& \frac{1}{1-q^2} \sum_{m=0}^\infty [m,0, \dots, 0]_n q^m
= \sum_{m_1=0}^\infty \sum_{m_2=0}^\infty [m_1,0, \dots, 0]_n q^{m_1+2m_2} 
\nonumber \\
&=&1+[1,0, \ldots, 0]_n q + \left( [2,0, \ldots, 0]_n
  +1\right)q^2+
\ldots 
\eea 
This suggests that this function can be
factorized from the expression for $Z_m$ and may make the expansion in
$q$ easier to compute.  In fact, the function has a particularly
simple form as a product over all weights. We quote the result for
$SO(9)$, 
\bea &&{PE \left [ [1,0, 0, 0]_9 q \right]} = 
\\   &&
\frac{1}{(1-q y_1) \left(1-\frac{q y_2}{y_1}\right) \left(1-\frac{q
      y_3}{y_2}\right) \left(1-\frac{q y_4^2}{y_3}\right)
  \left(1-\frac{q y_3}{y_4^2}\right) \left(1-\frac{q y_2}{y_3}\right)
  \left(1-\frac{q y_1}{y_2}\right) \left(1-\frac{q}{y_1}\right)
  (1-q)}.  
\nonumber 
\eea  
%CC
Intuitively speaking, for calculational reasons, we are
factoring the contribution from nine oscillator modes at level one
although not all are physical in the light-cone and we will therefore
pay a price. 
%The geometric meaning of the function is as
%follows. It counts all possible monomial functions with positive
%powers that one can write down in $\BR^9$.  
To proceed, define 
\beq Z_m = PE \left [ [1,0, \ldots, 0]_n q \right]
Z_m^\prime, \eeq 
Again, this is done for ease of computation. Since
not all of the factored modes are physical, there can now be negative
signs in the remaining expression.  For completeness we write down the
expansion of $Z_m^\prime$ to seventh order 
\bea
Z_m^\prime &=& 1 + \left ([0, 0, 0, 1]_9 - 1\right ) q^2 + \left ([0, 1, 0, 0]_9 + [1, 0, 0, 0]_9 - [0, 0, 0, 1]_9\right ) q^3 \nonumber \\
&+& \left ([1, 0, 0, 1]_9 + [0, 0, 0, 1]_9 - [1, 0, 0, 0]_9\right )
q^4
\nonumber \\
&+& \left ([0, 0, 0, 2]_9 + [1, 1, 0, 0]_9 + [2, 0, 0, 0]_9 + 1\right ) q^5 \nonumber \\
&+& \left ([2, 0, 0, 1]_9 + [0, 1, 0, 1]_9 + [0, 0, 1, 0]_9 + [1, 0, 0, 1]_9 + [1, 0, 0, 0]_9 + 1\right ) q^6 \nonumber \\
&+& \left ([1, 0, 0, 2]_9 + [2, 1, 0, 0]_9 + [3, 0, 0, 0]_9 + [0, 1, 0, 1]_9 + [1, 0, 0, 1]_9 + [1, 0, 1, 0]_9 \right. \nonumber \\
&+& \left.[1, 1, 0, 0]_9 + [0, 0, 1, 0]_9 + [0, 1, 0, 0]_9 + [0, 0, 0,
  1]_9 + [1, 0, 0, 0]_9 - 1\right ) q^7
\nonumber \\
&+& O(q^8) 
\eea 
This expression is an improved version of equation
\eref{pertZm} in the sense that it contains less terms.
%CC
 One can now repeat this
calculational simplification with higher oscillator modes if one so desires.

\subsection{Bosonic string theory}
Before we turn to applications of the above results, we show how the same method applies to bosonic 
string theories in twenty-six dimensions.
 For the left-movers one obtains the partition
function:

% With the same formalism it is possible to write down the partition
% function for the open bosonic string.  The number of space dimensions
% is 25 and therefore the little groups for massless and massive
% particles are $SO(24)$ and $SO(25)$, respectively.  The mass level is
% shifted by one unit due to the zero point contribution.  All together,
% the partition function takes the form
\beq Z_{Bosonic} = \frac{1}{q}
PE \left [ \frac{q}{1-q} \left ([1,0,\ldots, 0]_{25} -1\right ) \right
]. \eeq
The character of the vector representation of $SO(25)$ can be taken to be
the following function of the
fugacities $x_1,\ldots, x_{12}$: \bea
\left [1,0,\ldots, 0 \right ]_{25} = \sum_\alpha x^\alpha &=& x_1 + \frac{x_2}{x_1} + \frac{x_3}{x_2} + \ldots + \frac{x_{11}}{x_{10}} + \frac{x_{12}^2}{x_{11}} \nonumber \\
&+& \frac{1}{x_1} + \frac{x_1}{x_2} + \frac{x_2}{x_3} + \ldots +
\frac{x_{10}}{x_{11}} + \frac{x_{11}}{x_{12}^2} + 1, \eea where
$\alpha$ runs over the set of weights of the representation and
$x^\alpha$ is a multi-index notation.
The first few terms in the expansion are
\bea
Z_{Bosonic} &=& \frac{1}{q} + [1,0,\ldots,0]_{24}+ [2,0,\ldots,0]_{25}q
\nonumber \\
&+& ([3,0,\ldots,0]_{25}+[0,1,0,\ldots,0]_{25})q^2
\nonumber
\\
&+&([4,0,\ldots,0]_{25}+[2,0,\ldots,0]_{25}+[1,1,0,\ldots,0]_{25}+1)q^3
\nonumber \\ 
&+&([5,0,\ldots,0]_{25}+[3,0,\ldots,0]_{25}+[2,1,0,\ldots,0]_{25}
\\
\nonumber 
&& +[1,1,0,\ldots,0]_{25}+[1,0,\ldots,0]_{25}+[0,1,0,\ldots,0]_{25})q^4 + O(q^5).
\eea
We can easily  combine chiral halves and level match to obtain the closed bosonic string spectrum in
twenty-six dimensions.

\subsection{Superstring Theories}
To write down the partition function for the superstring theories, we need to expand
the chiral partition function \eref{ZOpen1} in powers of $q$,
\beq
Z_{Left} (q; z_1, z_2, z_3, z_4) = \sum_{n=0}^\infty d_n(z_1, z_2, z_3, z_4) q^n ,
\label{ZOpen2}
\eeq
where $d_n$ for $n>1$ is a sum of fermionic and bosonic representations of $SO(9)$.
The first few terms can be read from equation \eref{pertZm}
\bea
\nonumber
d_0 &=& [1,0,0,0]_8+[0,0,0,1]_8 \\
\nonumber
d_1 &=& [2,0,0,0]_9+[1,0,0,1]_9+[0,0,1,0]_9 \\
d_2 &=& [1,0,0,0]_9 d_1 
\eea
etcetera.

\subsubsection{Type II}
It should be clear that the generating function of Type II string
theories can be written as the product of $SO(9)$ covariant
left-moving and right-moving partition functions. To obtain
the physical spectrum it is sufficient to implement level
matching.
The massive spectrum at level $n$ is given by the tensor product of $d_n$'s.
\beq
\sum_{n=1}^\infty d_n( z_1, z_2, z_3, z_4) \tilde{d}_n(z_1,z_2,z_3,z_4) q^n \bar{q}^{n} .
\eeq
For Type IIB string theory, we have $\tilde{d}_n = d_n$ while for Type IIA
we switch the chirality of the zero-modes on the right:
 $\tilde{d}_0=[1,0,0,0]_8 + [0,0,1,0]_8$ and $\tilde{d}_n=d_n$ for
$n \ge 1$.
\subsubsection{Type I}
At one loop in Type I string theory, we have contributions from the torus,
the Klein bottle, the annulus and the M\"obius strip.
The contributions from the torus and Klein bottle take the form:
\beq
Z_{T+KB} (q; z_1, z_2, z_3, z_4) = \sum_{n=0}^\infty [d_n(z_1, z_2, z_3, z_4)]^2_{S,A} q^n \bar{q}^n,
\label{ZOpen3}
\eeq where the subscript $S,A$ mean that we symmetrize bosonic
representations in the chiral partition function $Z_{Left}$ while
we anti-symmetrize fermions in $Z_{Left}$. The net effect for space-time
bosons is to pick up
the symmetric part in the NS-NS sector and the
anti-symmetric part  of the R-R sector of Type IIB string theory.
Note that we already level-matched the spectrum.
The contribution from the annulus and the M\"obius
strip %requires an explicit expression for the character of the vector representation of $SO(32)$,
take the form \bea
\label{ZOpen4}
Z_{A+M} (q; z_1, z_2, z_3, z_4;s_1, \ldots, s_{16}) &=& [0,1,0,\ldots,0]_{32}
\sum_{n=0}^\infty d_{2n}(z_1, z_2, z_3, z_4) q^{2n}
\\ \nonumber
&+& ([2,0,\ldots,0]_{32}+1)\sum_{n=0}^\infty d_{2n+1}(z_1, z_2, z_3, z_4) q^{2n+1}.
\eea
We have taken the opportunity to also introduce fugacities that indicate the $SO(32)$ representation
content of the Chan-Paton degrees of freedom associated to the open strings.
The Type I partition function is the sum
\beq
Z_{Type I} (q; z_1, z_2, z_3, z_4;s_1, \ldots, s_{16}) = Z_{T+KB} 
%(q; z_1, z_2, z_3, z_4) 
+ Z_{A+M}
% (q; z_1, z_2, z_3, z_4;s_1, \ldots, s_{16})
.
\label{ZOpen5}
\eeq
Combining these equations, we find the familiar massless sector:
\bea
&&Z_{0, Type I} = [2,0,0,0]_8+[1,0,0,1]_8+[0, 1, 0, 0]_8 + [0, 0, 1, 0]_8 + [0, 0, 0, 0]_8 \nonumber \\
&+&[0,1,0,\ldots,0]_{32}([1,0,0,0]_8+[0,0,0,1]_8).
\label{ZOpen6}
\eea
We define the massive partition function:
\beq Z_{m,I}(q; y_1, y_2, y_3, y_4) = \frac{Z_{Type I} - Z_{0, Type I}}{Z_Q},
\eeq
where $Z_Q$ is the supermultiplet given in equation \eref{ZQ}.
The massive partition function $Z_{m,I}$ has the expansion:
\bea
Z_{m,I} &=& q( [2,0,\ldots,0]_{32} +1+[0,1,0,\ldots,0]_{32}[1, 0, 0, 0]_9 q +\dots) 
\nonumber \\
&+& q \bar{q} ([2,0,0,0]_9+[0, 0, 1, 0]_9\nonumber \\
&+&([4, 0, 0, 0]_9 + [2, 1, 0, 0]_9 + [2, 0, 1, 0]_9 
+ [2, 0, 0, 1]_9 + 2 [2, 0, 0, 0]_9 + [1, 1, 0, 1]_9
\nonumber\\
&+& [1, 1, 0, 0]_9 + [1, 0, 0, 2]_9 +  2 [1, 0, 0, 1]_9 
+ [0, 2, 0, 0]_9 + [0, 1, 0, 1]_9 + [0, 1, 0, 0]_9
\nonumber\\
&+& [0, 0, 1, 1]_9 + 2 [0, 0, 1, 0]_9 + [0, 0, 0, 1]_9 + [0, 0, 0, 0]_9) q \bar{q}
+ \dots).
\nonumber\\
\label{ZOpen7}
\eea
The power of $q$ in the open string contributions
is equal to the mass squared in string units, $\alpha' m^2$,
while the power of $q \bar{q}$ in the closed string contributions is
$\alpha' m^2 /2$.
\subsubsection{Heterotic $SO(32)$}
Let's discuss the $SO(32)$ heterotic string theory next. Again, we have a $SO(32)$ gauge group, but including
the fugacities that code the gauge group representation content of all excitations is now slightly more involved.
 The charged sector is generated by thirty-two fermionic generators and hence we define the fermionic plethystic exponential
\beq
Z_{F,32} (q; f; s_1, \ldots, s_{16}) = PE_F \left[ \frac{f}{1-q} [1,0,\ldots,0]_{32} \right],
\label{ZF,32)}
\eeq
where $s_1,\ldots, s_{16}$ are the 16 fugacities of $SO(32)$, and below only the $f$ dependence is explicit, namely $Z_{F,32}(f)$.
The NS contribution is as above,
\beq Z_{NS,32} =
\frac{1}{2 q}\left(Z_{F,32}\left(\sqrt{q}\right)+Z_{F,32}\left(-\sqrt{q}\right)\right),
\label{ZNSh}
\eeq
and similarly the R sector gets 2 contributions from
\beq Z_{32,\pm} =
\frac{q}{2}\left(Z_{F,32}\left({q}\right)\pm Z_{F,32}\left(-{q}\right)\right).
\label{Zpmh}
\eeq
The RNS contribution takes the form  
\beq Z_{RNS,32}(q; s_1, \ldots, s_{16}) = 
%\frac{Z_{F,32}\left(\sqrt{q}\right)-Z_{F,32}\left(-\sqrt{q}\right)}{2q}
Z_{NS,32}+[0,\ldots,0,1]_{32} Z_{32,+} + [0,\ldots,0,1,0]_{32} Z_{32,-}.  \eeq
We further need the contribution from the eight bosonic oscillators, as in equation \eref{ZB}, to construct the right moving sector
\beq
Z_{Right,32} (q; z_1, z_2, z_3, z_4; s_1,\ldots, s_{16}) = Z_B Z_{RNS,32} .
\eeq
The first terms read
\bea
\label{massSO32}
&&Z_{Right,32} (q; z_1, z_2, z_3, z_4; s_1,\ldots, s_{16}) =
\frac{1}{q}+[1,0,0,0]_8+[0,1,0,\ldots,0]_{32} \nonumber \\ \nonumber &+& ([2, 0,
0, 0]_9 + [1, 0, 0, 0]_9 [0, 1, 0,\ldots,0]_{32} + [2,
0,\ldots,0]_{32} 
\\
\nonumber 
&+& [0, 0, 0,1, 0,\ldots,0]_{32} + [0, \ldots,0,
1]_{32} + 1) q \\ \nonumber &+&([3, 0, 0, 0]_{9} + [2, 0, 0, 0]_{9}
[0, 1, 0,\ldots, 0]_{32} + [0, 1, 0, 0]_{9} \\ \nonumber &+& [1, 0, 0,
0]_{9} ([2, 0,\ldots, 0]_{32} + [0, 1,0,\ldots, 0]_{32} + [0, 0, 0,
1,0\ldots,0]_{32} + [0, \ldots,0, 1]_{32}+1) \\ \nonumber &+& [1, 0,
1, 0,\ldots,0]_{32} + [1, 0,\ldots,0, 1,0]_{32} + 2 [0, 1, 0,\ldots,
0]_{32} + [0, 0, 0, 0, 0, 1,0,\ldots,0]_{32}) q^2 \\ \nonumber &+&([4,
0, 0, 0]_9 + [3, 0, 0, 0]_9 [0, 1, 0, \ldots, 0]_{32} + [1, 1, 0, 0]_9
\\ \nonumber &+& [2, 0, 0, 0]_9 ([2, 0,\ldots, 0]_{32} + [0, 1,
0,\ldots, 0]_{32} + [0, 0, 0, 1, 0,\ldots, 0]_{32} + [0,\ldots, 0,
1]_{32} + 2) \\ \nonumber &+& [1, 0, 0, 0]_9 ([2, 0,\ldots, 0]_{32} +
[1, 0, 1, 0,\ldots, 0]_{32} + [1, 0,\ldots, 0, 1,0]_{32} + 3 [0, 1,
0,\ldots, 0]_{32} \\ \nonumber &+& [0, 0, 0, 1, 0,\ldots, 0]_{32} +
[0, 0, 0, 0, 0, 1, 0,\ldots, 0]_{32} + [0,\ldots, 0, 1]_{32} + 1) \\
\nonumber &+& [0, 1, 0, 0]_9 [0, 1, 0,\ldots, 0]_{32} + 2 [2,
0,\ldots, 0]_{32} + [1, 0, 1, 0,\ldots, 0]_{32} + [1, 0, 0, 0, 1,
0,\ldots, 0]_{32} \\ \nonumber &+& [1, 0,\ldots, 0, 1, 0]_{32} + [0,
2, 0,\ldots, 0]_{32} + [0, 1, 0,\ldots, 0, 1]_{32} 
\\
\nonumber 
&+& [0, 1, 0,\ldots,
0]_{32} + 2 [0, 0, 0, 1, 0,\ldots, 0]_{32} \\  &+& [0, 0, 0,
0, 0, 0, 0, 1, 0,\ldots, 0]_{32} + [0,\ldots, 0, 1]_{32} +
3)q^3+O(q^4).  \eea This expression satisfies the well known condition
which states that only two out of the four conjugacy classes of $SO(32)$
are present in the perturbative spectrum - the adjoint class and the spinor
class. In terms of the sixteen Dynkin labels this condition means that the
sum of odd entries should be 0 mod 2. Indeed one can check that all
representations of $SO(32)$ in \eref{massSO32} satisfy this condition.

To get the final answer, one combines the right-moving partition function
with the left-moving super-symmetric partition function and level-matches.

%By setting all non-Abelian fugacities to 1 we find the expansion \beq
%Z_{right,32} (q; 1, 1, 1, 1; 1,\ldots, 1) = \frac{1}{q}+504+73764 q +
%2695040 q^2 + 54755730 q^3 + O(q^4), \eeq%%
%
%a formula which is good also for the $E_8\times E_8$ string.
\subsubsection{Heterotic $E_8 \times E_8 $}
For the $E_8\times E_8$ theory, we proceed similarly.
Define the fermionic plethystic exponential for $SO(16)$
\beq
Z_{F,16} (q; f; s_1, \ldots, s_{8}) = PE_F \left[ \frac{f}{1-q} [1,0,\ldots,0]_{16} \right],
\label{ZF,16)}
\eeq
where $s_1,\ldots, s_{8}$ are the 8 fugacities of $SO(16)$, and below only the $f$ dependence is explicit, $Z_{F,16}(f)$.
The NS contribution is
\beq Z_{NS,16} = \frac{Z_{F,16}\left(\sqrt{q}\right)+Z_{F,16}\left(-\sqrt{q}\right)}{2 \sqrt{q}},
\label{ZNS16}
\eeq
and the R sector gets two contributions from
\beq Z_{16,\pm} =
\frac{\sqrt{q}}{2}\left(Z_{F,16}\left({q}\right)\pm Z_{F,16}\left(-{q}\right)\right).
\label{Zpm16}
\eeq
The RNS contribution takes the form
\beq Z_{RNS,16}(q; s_1, \ldots, s_{8})
 = 
%\frac{Z_{F,16}\left(\sqrt{q}\right)-Z_{F,16}\left(-\sqrt{q}\right)}{2\sqrt{q}}
Z_{NS,16} +[0,\ldots,0,1]_{16} Z_{16,+} + [0,\ldots,0,1,0]_{16} Z_{16,-}.  
%\nonumber
\eeq
Collecting all contributions, including the 8 bosonic oscillators, taken from equation \eref{ZB} the right moving sector becomes
\beq
Z_{Right,E8\times E8} (q; z_1, z_2, z_3, z_4; s_1,\ldots, s_{16}) = Z_B Z_{RNS,16}(q,s_1,\ldots,s_8)Z_{RNS,16}(q,s_9,\ldots,s_{16})
\nonumber .
\eeq
The first few terms read
\bea
&&Z_{Right,E8\times E8} (q; z_1, z_2, z_3, z_4; s_1,\ldots, s_{16}) = \frac{1}{q}
\\ \nonumber
&+&[1,0,0,0]_8+[0,0,0,0,0,0,0,1]_{E8_1}+[0,0,0,0,0,0,0,1]_{E8_2} \\ \nonumber
&+& ([2, 0, 0, 0]_9 + [1, 0, 0, 0]_9 \left( [0,0,0,0,0,0,0,1]_{E8_1}+[0,0,0,0,0,0,0,1]_{E8_2} \right)
\\ \nonumber
&+&[1, 0,0,0,0,0,0,0]_{E8_1} +[1,0,0,0,0,0,0,0]_{E8_2} 
\\ \nonumber
&+& 
[0,0,0,0,0,0,0,1]_{E8_1}[0,0,0,0,0,0,0,1]_{E8_2} + 2) q
\\ \nonumber
&+&([3, 0, 0, 0]_9 +[0, 1, 0, 0]_9 + 2[1, 0, 0, 0]_9 
\\ \nonumber
&+& [2, 0, 0, 0]_9 ([0,0,0,0,0,0,0,1]_{E8_1}+[0,0,0,0,0,0,0,1]_{E8_2})
\\ \nonumber
&+&[1, 0, 0, 0]_9 ([1,0,\dots,0]_{E8_1} +[1,0,\dots,0]_{E8_2} +[0,\dots,0,1]_{E8_1} +[0,0,0,0,0,0,0,1]_{E8_2}
\\ \nonumber
&+& [0,0,0,0,0,0,0,1]_{E8_1}[0,0,0,0,0,0,0,1]_{E8_2}) + [1,0,0,0,0,0,0,0]_{E8_1}[0,0,0,0,0,0,0,1]_{E8_2}
\\ \nonumber
&+& [0,0,0,0,0,0,0,1]_{E8_1} [1,0,0,0,0,0,0,0]_{E8_2} + [0,0,0,0,0,0,0,1]_{E8_1}[0,0,0,0,0,0,0,1]_{E8_2}
\\ \nonumber
&+& [0,0,0,0,0,0,1,0]_{E8_1}+[0,0,0,0,0,0,1,0]_{E8_2}
+ 2 [0,\dots,0,1]_{E8_1}+2[0,\dots,0,1]_{E8_2}) q^2
\\ \nonumber
&+&O(q^3).
\eea
For reference we record the dimensions of the E8 representations which enter into this formula,
\bea
dim[0,0,0,0,0,0,0,1]_{E8} &=& 248,
\\ \nonumber
dim[1,0,0,0,0,0,0,0]_{E8} &=& 3875,
\\ \nonumber
dim[0,0,0,0,0,0,1,0]_{E8} &=& 30380.
\\ \nonumber
\eea
It is possible to treat the $E_8 \times E_8$ quantum numbers covariantly from
the start by bosonizing the fermions, and using that the momentum lattice of the
bosons is the $E_8 \times E_8$ root lattice. 

\section{Conclusions and some open problems}
We gave the $SO(D-1)$ covariant form of massive string spectra in 
$\mathbb{R}^{D-1,1}$ for various string theories.
Our method generalizes to other Minkowski compactifications of string
theory. We already illustrated how one can generate further generalizations
of partition functions that also code the gauge group representation content
at all mass levels. Our method is also applicable to compactifications with isometries, or
super-isometries
where we may choose to introduce fugacities for the Cartan subalgebra
of the super-isometry group. Generating functions that code the 
representation content of compactificications
with discrete symmetries (like Calabi-Yau compactifications at Gepner
points) can also be written down. These are fairly straightforward
generalizations of our results, and the resulting counting functions
are likely to have interesting (modular and other) properties.

It is a further challenge to generalize these counting functions to
other symmetric string backgrounds, including $AdS$ space-times.

%%%%%%%%%%%%%%%%%%%%%%%%%%%%

\section*{Acknowledgments}
 We particularly thank Costas Bachas for very useful
discussions at the early stage of the project.  Discussions with Atish
Dabholkar, Vishnu Jejjala, Giuseppe Policastro, Ashoke Sen and Alberto
Zaffaroni are also greatly appreciated, as well as correspondence with
Massimo Bianchi, Michael Duff, Pierre Ramond and John Schwarz. We thank
Andrew Thomson for pointing out various typographical mistakes in our
paper.   A.~H.~
would moreover like to thank the \'Ecole Normale Sup\'erieure and the \'Ecole
Polytechnique, Paris as well as the YITP, Kyoto, for their kind
hospitality during the various stages of this work.  This research
project is supported in part by ERC Advanced Grant 226371 and the
grant ANR-09-BLAN-0157-02.

\end{document}